\documentstyle[12pt,epsfig]{article}
\newlength{\dinwidth}
\newlength{\dinmargin}
\newcommand{\ba}{\begin{array}}
\newcommand{\ea}{\end{array}}
\newcommand{\beq}{\begin{equation}}
\newcommand{\eeq}{\end{equation}}
\newcommand{\bea}{\begin{eqnarray}}
\newcommand{\eea}{\end{eqnarray}}

\def\S{{\bf S}}

\def\bce{\begin{center}}
\def\ece{\end{center}}

\def\nonu{\nonumber}

\def\pa{\partial}
\def\al{\alpha}
\def\be{\beta}
\def\ga{\gamma}

\def\De{\Delta}

\def\th{\theta}

\def\si{\sigma}

\def\R{{\bf R}}

\def\S{{\bf S}}

\begin{document}
\thispagestyle{empty}
\addtocounter{page}{-1}
\begin{flushright}
{\tt hep-th/0205109}\\
\end{flushright}
\vspace*{1.3cm}
\centerline{\Large \bf Comments on Penrose Limit of
$AdS_4 \times M^{1,1,1}$}
\vspace*{1.5cm} 
\centerline{ \bf Changhyun Ahn
}
\vspace*{1.0cm}
\centerline{\it 
Department of Physics,}
\vskip0.3cm 
\centerline{  \it Kyungpook National University,}
\vskip0.3cm
\centerline{  \it  Taegu 702-701, Korea }
\vspace*{0.3cm}
\centerline{\tt ahn@knu.ac.kr 
}
\vskip2cm
\centerline{\bf  abstract}
\vspace*{0.5cm}

We construct a Penrose limit of $AdS_4 \times M^{1,1,1}$
where $M^{1,1,1}=\frac{
SU(3) \times SU(2) \times U(1)}{SU(2) \times U(1) 
\times U(1)}$
that provides the pp-wave geometry equal to the one in the Penrose limit
of $AdS_4 \times \S^7$. 
There exists a subsector of 
three dimensional ${\cal N}=2$ dual gauge theory which 
has enhanced ${\cal N}=8$ maximal supersymmetry.
We identify operators in the ${\cal N}=2$ gauge theory with
supergravity KK excitations in the pp-wave geometry and 
describe how the gauge theory operators made out of
two kinds of chiral fields of conformal dimension 4/9, 1/3 fall into 
${\cal N}=8$ supermultiplets.

\vspace*{4.0cm}

%\begin{flushleft}
%{Aug., 1999}\\
%\end{flushleft}
%\centerline{\bf DRAFT VERSION}
%\centerline{Submitted to Nuclear Physics B}
\baselineskip=18pt
\newpage

%%%%%%%%%%%%%%%%%%%%%%%%%%%%%%%%%%%%%%%%%%%%%%%%%%%%%%%%%%%%%%%%%%%%%%%%%%%%%
\section{Introduction}
%%%%%%%%%%%%%%%%%%%%%%%%%%%%%%%%%%%%%%%%%%%%%%%%%%%%%%%%%%%%%%%%%%%%%%%%%%%%%

Recently \cite{bmn}, it was found that the large $N$ limit of a subsector of
four dimensional ${\cal N}=4$ $SU(N)$ supersymmetric gauge theory is dual to
type IIB string theory in the pp-wave background \cite{bfhpetal,bfhpetal1}.
In the $N \rightarrow \infty$ with the finiteness of string coupling
constant, this subspace of the gauge theory and the operator algebra are
described by string theory in the pp-wave geometry.
By considering a scale limit of the geometry near a null geodesic in $AdS_5 
\times \S^5$, it leads to the appropriate subspace of the gauge 
theory. The operators in the subsector of ${\cal N}=4$ gauge theory can be 
identified with the excited states in the pp-wave background. 
There are many papers \cite{mt}-\cite{constableetal} 
related to the work of \cite{bmn}.
There exist some works \cite{ikm,go,zs}
on the Penrose limit of $AdS_5 \times T^{1,1}$ that
gives the pp-wave geometry of $AdS_5 \times \S^5$. 
There is a subsector of ${\cal N}=1$ gauge theory
that contains an enhanced ${\cal N}=4$ supersymmetry. The corresponding 
operators in the gauge theory side were identified with the stringy 
excitations in the pp-wave geometry and some of the gauge theory operators
are combined into ${\cal N}=4$ supersymmetry multiplets \cite{go}.  
Moreover, it was found in \cite{ahn02} that a subsector of
$d=3, {\cal N}=2$ gauge theory dual to $AdS_4 \times Q^{1,1,1}$ has
enhanced ${\cal N}=8$ maximal supersymmetry and the gauge theory operators
are combined into ${\cal N}=8$ chiral multiplets. 

In this paper, we consider a similar duality
that is present between a certain three dimensional ${\cal N}=2$
gauge theory and 11-dimensional supergravity theory
in a pp-wave background with the same spirit as in \cite{ahn02,ikm,go,zs}. 
We describe this duality by taking a scaling limit of the duality
between 11-dimensional supergravity on $AdS_4 \times M^{1,1,1}$
where $M^{1,1,1}$ was considered first in \cite{witten} 
and three dimensional superconformal field theory
that consists of an ${\cal N}=2$ 
$SU(N) \times SU(N) $ gauge theory
with two kinds of chiral fields $U_i, i=1,2,3$ transforming in the 
$\left( \Box\!\Box, {\Box\!\Box}^{\ast} \right) $  
color representation and 
$V_A, A=1,2$ transforming in the 
$\left( {\Box\!\Box\!\Box}^\ast, {\Box\!\Box\!\Box} \right) $  
color representation  
\cite{fabbrietal}.
The complete analysis on the spectrum of
$AdS_4 \times M^{1,1,1}$ was found by \cite{fabbri1}. 
This gives the theory of \cite{fabbrietal} that lives on $N$ M2-branes 
at the conical singularity of a Calabi-Yau four-fold. 
The scaling limit is obtained by considering the geometry near a null 
geodesic carrying large angular momentum in the $U(1)_R$
isometry of the $M^{1,1,1}$ space which is dual to the $U(1)_R$ R-symmetry
in the ${\cal N}=2$ superconformal field theory.
In section 2, we construct the scaling limit around
a null geodesic in $AdS_4 \times M^{1,1,1}$ and obtain a pp-wave 
background.
In section 3, we identify supergravity KK excitations obtained by
\cite{fabbri1} in the Penrose limit
with gauge theory operators.
In section 4, we summarize our results.

%%%%%%%%%%%%%%%%%%%%%%%%%%%%%%%%%%%%%%%%%%%%%%%%%%%%%%%%%%%%%%%%%%%%%%%%%%%%%
\section{Penrose Limit of $AdS_4 \times M^{1,1,1}$}
%%%%%%%%%%%%%%%%%%%%%%%%%%%%%%%%%%%%%%%%%%%%%%%%%%%%%%%%%%%%%%%%%%%%%%%%%%%%%

Let us start with the supergravity solution dual to
the ${\cal N}=2$ superconformal field theory \cite{fabbrietal}.
By putting a large number of $N$ coincident M2 branes at the
conifold singularity and taking the near horizon limit,
the metric becomes  that \cite{pp1,np,pope} of 
$AdS_4 \times M^{1,1,1}$(See also \cite{cdf,df})
\bea
ds_{11}^2 =ds_{AdS_4}^2 +ds_{M^{1,1,1}}^2,
\label{metric}
\eea
where
\bea
ds_{AdS_4}^2 & = & L^2 \left( -\cosh^2 \rho \; d t^2 +d \rho^2 + 
\sinh^2 \rho \; d \Omega_2^2 \right),
\nonu \\
ds_{M^{1,1,1}}^2 & = & 
\frac{L^2}{64}  \left( d\tau + 3 \sin^2 \mu \sigma_3 + 2
\cos \theta d \phi \right)^2 +
\frac{3L^2}{4} \left( d \mu^2 + \frac{1}{4} \sin^2 \mu 
\left( \sigma_1^2 + \sigma_2^2 +\cos^2 \mu \sigma_3^2 \right) \right) \nonu \\
& &
 + \frac{L^2}{8}   \left( d \th^2+\sin^2 \th d \phi^2
\right),
\nonu
\eea
where $ d \Omega_2$ is the volume form of a unit $\S^2$
and the curvature radius 
$L$ of $AdS_4$ is given by
$(2L)^6= 32 \pi^2 \ell_p^6 N$. 
Topologically $M^{1,1,1}$ \footnote{
Sometimes instead of using the notation $M^{pqr}$,
$M(m,n)$ space with two parameters $m$ and $n$ 
(that are two winding numbers of the $U(1)$ gauge field
over the ${\bf CP}^2$ and $\S^2$ of the base manifold) rather than
three parameters $p, q,$ and $r$ in $M^{pqr}$ 
is used where the parameters are related by $m/n=3p/(2q)$. 
The integers $p,q$ and $r$ characterize the embedding of
$SU(2) \times U(1) \times U(1)$ in $SU(3) \times SU(2) \times U(1)$.
For $M(3,2)=M^{1,1,1}$, there is an ${\cal N}=2$ supersymmetry while
for all other $M(m,n)$ no supersymmetry survives.
In particular, $M(0,1)=M^{0,1,0}= {\bf CP}^2 \times \S^3$,
$M(1,0)= \S^5 \times \S^2$ and 
$M^{1,0,1}=(\S^5/{\bf Z}_3) \times \S^2$.
} 
is a nontrivial $U(1)$ bundle
over ${\bf CP}^2 \times \S^2$. 
The spherical coordinates $(\theta, \phi)$ parametrize
two sphere, as usual, and the angle $\mu$ and three real one forms
$\si_1, \si_2$ and $\si_3$ parametrize ${ \bf CP}^2$ satisfying the $SU(2)$
algebra and
the angle $\tau$ parametrizes the $U(1)$ Hopf 
fiber.
The angles vary over the ranges, $0 \leq \theta \leq  \pi,
0 \leq \phi \leq 2 \pi, 0 \leq \tau \leq 4 \pi$ and $ 0 \leq \mu 
\leq \pi/2$.
The $SU(3)  \times SU(2) \times U(1) $ isometry
 group of $M^{1,1,1}$ consists of $SU(3) \times SU(2) 
 $ global symmetry and $U(1)_R$ symmetry of the dual 
superconformal field theory of \cite{fabbrietal}.

Let us make a scaling limit around a null geodesic in 
 $AdS_4 \times M^{1,1,1}$ that rotates along the $\tau$ coordinate
of $M^{1,1,1}$ whose shift symmetry corresponds to
the $U(1)_R$ symmetry in the dual superconformal field theory.  
Let us introduce
coordinates which label the geodesic   
\bea
x^{+}  =  \frac{1}{2} \left( t +\frac{1}{8} \left( \tau + 
2 \phi \right) \right), \qquad
x^{-}  =   \frac{L^2}{2} \left( t -\frac{1}{8} \left( \tau + 2 \phi
\right) \right),
\nonu
\eea
and make a scaling limit around $\rho=0=\mu=\theta$
in the above geometry (\ref{metric}).
By taking the limit $L \rightarrow \infty$ while 
rescaling the coordinates 
\bea
\rho=\frac{r}{L}, \qquad \mu =\frac{ \zeta}{\sqrt{3} L}, \qquad
  \theta =\frac{\sqrt{2} \xi}{L}, \nonu
\eea
the Penrose limit of the  $AdS_4 \times M^{1,1,1}$ becomes 
\bea
ds_{11}^2 & = & -4 dx^{+} dx^{-} +
 \sum_{i=1}^3 \left( dr^i dr^i -r^i r^i dx^{+}
dx^{+} \right) \nonu \\
&& + \frac{1}{4}
\left( d\xi^2 +\xi^2 d\phi^2
- 2 \xi^2 d\phi dx^{+} \right) + \frac{1}{4} \left(
d\zeta^2 +  \zeta^2 \left( \widetilde{\si_1}^2 +
\widetilde{\si_2}^2+ \widetilde{\si_3}^2 \right) + 2 \zeta^2 
\widetilde{\si_3} dx^{+} \right)
\nonu \\
&= &  -4 dx^{+} dx^{-} +\sum_{i=1}^3 \left( dr^i dr^i -r^i r^i dx^{+}
dx^{+} \right) \nonu \\
& & + \frac{1}{4}
 \left( d w d \bar{w} 
+  i  \left( \bar{w} d w - w d \bar{w} \right) dx^{+} \right)
+ \frac{1}{4}
\sum_{i=1}^2 \left( d z_i d \bar{z}_i 
-  i  \left( \bar{z}_i d z_i - z_i d \bar{z}_i \right) dx^{+} \right)
\label{ppwave}
\eea
where in the last line we introduce the complex coordinate $w = \xi 
e^{i \phi}$ for $\R^2$ 
and a pair of complex coordinates $z_1$ and $z_2$  for $\R^4$. 
Since the metric has a covariantly constant null Killing
vector $\pa / \pa_{x^{-}}$, it is also pp-wave metric.
The pp-wave has a decomposition of the 
$\R^9$ transverse space into $\R^3 \times \R^2 \times \R^4 $ where
$\R^3$ is parametrized by $r^i$, $\R^2$ by $w$ and $\R^4$ by $z_1$ and 
$z_2$.
The symmetries of this background are the $SO(3)$ rotations in $\R^3$. 
%and
%a $U(1)_1 \times U(1)_2 \times U(1)_3$ symmetry corresponding to 
%the rotations of
%$\R^2 \times \R^2 \times \R^2$. 
In the gauge theory side,  
the $SO(3)$ symmetry corresponds to the subgroup of
the $SO(2,3)$ conformal group.
% and $U(1)_1 \times U(1)_2 \times U(1)_3$  
%rotation 
%charges $J_1, J_2$ and $J_3$ correspond to the linear combination of
%$Q_1$ and $R$, $Q_2$ and $R$, and $Q_3$ and $R$
%respectively where $R$ is the $U(1)_R$ charge of the gauge theory and $Q_1,
%Q_2$ and $Q_3$ are the Cartan generators of the $SU(2)_1 \times SU(2)_2 \times
%SU(2)_3$ global symmetry of the dual superconformal field theory.  
Note that the pp-wave geometry (\ref{ppwave}) in the scaling limit reduces to
the maximally ${\cal N}=8$ supersymmetric pp-wave solution of
$AdS_4 \times \S^7$ \cite{kow,op,gns,fk} through
$w = e^{i x^{+}} \widetilde{w}$ and $z_i = e^{-i x^{+}} \widetilde{z_i}$  
\bea
ds_{11}^2 = -4 d x^{+} d x^{-} +\sum_{i=1}^9  dr^i dr^i -
\left( \sum_{i=1}^3 r^i r^i  + \frac{1}{4}  \sum_{i=4}^9 
r^i r^i \right) dx^{+} dx^{+}. 
\nonu
\eea 
The supersymmetry enhancement in the Penrose limit
implies that a hidden ${\cal N}=8$ supersymmetry is present in the 
corresponding subsector of the dual ${\cal N}=2$ superconformal field 
theory.  In the next section, we provide precise description
of how to understand the excited states in the supergravity theory that 
corresponds in the dual superconformal field theory to operators with a
given conformal dimension. 

%%%%%%%%%%%%%%%%%%%%%%%%%%%%%%%%%%%%%%%%%%%%%%%%%%%%%%%%%%%%%%%%%%%%%%%%%%%%%
\section{Gauge Theory Spectrum}
%%%%%%%%%%%%%%%%%%%%%%%%%%%%%%%%%%%%%%%%%%%%%%%%%%%%%%%%%%%%%%%%%%%%%%%%%%%%%

The 11-dimensional supergravity theory in $ AdS_4 \times M^{1,1,1}$
is dual to the ${\cal N}=2$ gauge theory encoded in a quiver diagram
with gauge group
$SU(N) \times SU(N) $ 
with two kinds of chiral fields $U_i, i=1,2,3$ transforming in the 
$\left( \Box\!\Box,{\Box\!\Box}^\ast \right) $  color representation and 
$V_A, A=1,2$ transforming in the 
$\left( {\Box\!\Box\!\Box}^\ast, {\Box\!\Box\!\Box} \right) $  
color representation  
\cite{fabbrietal}.
At the fixed point, these chiral superfields $U, V$ have conformal 
weights $4/9, 1/3$ 
respectively \footnote{
There exist two supersymmetric 5-cycles that are the restrictions of the 
$U(1)$ fibration over ${\bf P}^{2 \ast}$ and ${\bf P}^1 \times {\bf P}^1$. One 
can determine the dimensions of the baryons by computing the ratio of 
volume of 5-cycle to the one of $M^{1,1,1}$. It turns out that
$N$ product of $U$ is equal to $4N/9$ and $N$ product of
$V$ is equal to $N/3$. Therefore this tells us that the conformal dimensions of
$U$ and $V$ are $4/9$ and $1/3$ respectively \cite{fabbrietal}. 
  } and
transform as $({\bf 3},{\bf 1})$ 
and $({\bf 1},{\bf 2})$ under the $SU(3) \times SU(2) 
$ global symmetry.
We identify states in the supergravity containing both short
and long multiplets with operators in the gauge theory
and focus only on the bosonic excitations of the theory.  
In each multiplet, we specify a $SU(3) \times SU(2)$ representation 
\footnote{A representation of $SU(3)$ can be identified by a Young diagram
and when we denote the Dynkin label $(M_1, M_2)$ so that totally
we have $M_1+2M_2$ boxes, the dimensionality of
an irreducible representation is $N(M_1,M_2)=(1+M_1)(1+M_2)
(\frac{2+M_1+M_2}{2})$.  
Also an irreducible representation of $SU(2)$
can be described by a Young diagram with $2J$ boxes. Its dimensionality 
is $2J+1$.  }, conformal
weight and $R$-charge.

$\bullet$ {\bf Massless(or ultrashort) multiplets} \cite{fabbri1}

$1)$ Massless graviton multiplet: 
$
({\bf 1}, {\bf 1}), \qquad
 \Delta=2, \qquad R=0.
$ 

There exists a stress-energy tensor superfield $T_{\al \be}(x, \theta)$ 
that satisfies
the equation for conserved current $D_{\al}^{\pm} T^{\al \be}(x, \theta) 
=0$. 
This $T_{\al \be}(x, \theta)$ 
is a singlet with respect to 
the flavor group $SU(3) \times SU(2) 
$ and its conformal dimension is 2. Moreover $R$ charge is 0.
So this corresponds to the massless graviton multiplet  that 
propagates in the $AdS_4$ bulk. 

$2)$ Massless vector multiplet:
$
({\bf 8}, {\bf 1}) \;\;\; \mbox{or}  \;\;\; ({\bf 1}, {\bf 3}), \qquad
 \Delta=1, \qquad R=0.
$
 
There exists a conserved vector current, a scalar superfield 
$J_{SU(3)}(x, \theta)$,
to the generator of the flavor
symmetry group  $SU(3)$ through Noether theorem satisfying the conservation
equations $D^{\pm \al} D^{\pm}_{ \al} J_{SU(3)}(x, \theta) = 
 0 $. 
This $J_{SU(3)}(x, \theta)$ transforms  in the adjoint  representation
$\bf 8$ of the first factor $SU(3)$ of 
the flavor group and its conformal dimension is 1 with vanishing
$R$-charge.
This corresponds to the massless vector multiplet $({\bf 8}, {\bf 1})$ 
propagating in the $AdS_4$ bulk. 
There is also massless vector multiplet denoted by
$({\bf 1}, {\bf 3})$. 
There exists a corresponding scalar superfield 
$J_{SU(2)}(x, \theta)$,
to the generator of the flavor
symmetry group  $SU(2)$ satisfying the conservation
equations $D^{\pm \al} D^{\pm}_{ \al} J_{SU(2)}(x, \theta) = 
 0 $. 
This $J_{SU(2)}(x, \theta)$ transforms  in the adjoint  representation
$\bf 3$ of the $SU(2)$ of 
the flavor group and its conformal dimension is 1 with vanishing
$R$-charge.

$3)$ Two massless vector multiplets:
$
({\bf 1}, {\bf 1}), \qquad
 \Delta=1, \qquad R=0.
$
 
It is known that  the Betti current $J_{\mbox{betti}} = 2 U \overline{U}-
3 V \overline{V}$ of $M^{1,1,1}$ is obtained from
the toric description \cite{fabbrietal}, 
it is  conserved and corresponds to
additional massless vector multiplets. Therefore massless multiplets 1), 2)
and 3) saturate
the unitary bound and have a conformal weight related to the 
maximal spin. 

$\bullet$ {\bf Short multiplets} \cite{fabbri1}

It is known that 
the dimension of the scalar operator in terms of energy labels,
in the dual SCFT corresponding
$AdS_4 \times M^{1,1,1}$ is
\bea
\Delta = \frac{3}{2} + \frac{1}{2} \sqrt{1 + \frac{m^2}{4}} =
\frac{3}{2} + \frac{1}{2} \sqrt{45 + \frac{E}{4} -6 \sqrt{36 +E}} .
\label{delta}
\eea 
The energy spectrum on $M^{1, 1, 1}$ exhibits an interesting feature
which is relevant to superconformal algebra and it is given by
\cite{df,fabbri1} \footnote{
Of course, for general $M^{pqr}$ space, it is known that 
the energy spectrum on this space is given by 
\bea
E = \ga^2 \left( \frac{2}{3} \al \frac{q^2}{p^2} \left(M_1 +M_2 +
M_1 M_2 \right) + 2\be \left( J +J^2 -\frac{1}{4} q^2 Y^2\right)
+\frac{1}{4} q^2 Y^2 \right) 
\nonu
\eea
where
$M_1 =0, 1, 2, \cdots, M_2 =M_1 +\frac{3}{2} p Y, J=|\frac{1}{2} qY|, 
|\frac{1}{2}q Y|+1, \cdots,$ and $ Y =0, \pm 2, \pm 4, \cdots$. For
$Y<0$, $M_2 =0,1,\cdots$ and $M_1 =M_2 +\frac{3}{2} p |Y|$. 
Here $\al, \be$ and $\ga$ are related to the scale parameters.
In particular,
on the supersymmetric space $M^{1,1,1}$, we have $p=q=1, \al =\frac{1}{2},
\be =\frac{1}{4}$ and $\ga=8$. }
\bea
E =  64 
\left( \frac{1}{3} \left( M_1 +M_2 + M_1 M_2 \right) +
\frac{1}{2} J(J+1) + \frac{1}{8} Y^2 \right)
\label{energy}
\eea
where the eigenvalue $E$ is classified by
$SU(3)$ quantum numbers $(M_1, M_2)$,
$SU(2)$ isospin $J$ and
weak hypercharge $Y$: $M_1=0,1,2, \cdots$, $M_2=M_1 +\frac{3Y}{2}$,
$J=|Y/2|, |Y/2|+1, \cdots$ and $Y=0, \pm 2, \pm 4, \cdots$. For 
$Y<0$, $M_2=0,1,2, \cdots$ and $M_1=M_2+\frac{3|Y|}{2}$. 

The corresponding eigenmodes occur in $( M_1, M_1+\frac{3|Y|}{2})$ for
$Y>0$ or, $( M_2+\frac{3|Y|}{2}, M_2)$ for
$Y<0$ $SU(3)$ representation,   
the angular momentum $J$ $SU(2)$ representation and 
$U(1)$ charge $Y/4$.
The eigenvalues (\ref{energy})
as a linear combination of the quadratic Casimirs for the symmetry group
$SU(3) \times SU(2) \times U(1)$ are the form for a coset manifold \cite{pope}
sometime ago.

The $U(1)$ part of the isometry goup of $M^{1, 1, 1}$
acts by shifting $U(1)$ weak hypercharge $Y$. 
The half-integer $R$-charge, $R$ is related to $U(1)$ charge $Y$
by $R=Y$. 
Let us take $R \geq 0$. One can do similar case for $R < 0$.
  One can find the lowest value of $\Delta$ is equal to
$R$ corresponding to a mode scalar with $M_1=0, M_2=3R/2$ and $J=R/2$ 
because $E$ 
becomes $16R(R+3)$ and plugging back to (\ref{delta})
then one obtains $\De=R$. 

Thus we find
a set of operators filling out a ${\bf ( \frac{(1+
\frac{3R}{2})(2+\frac{3R}{2})}{2}, R+1)}$ 
multiplet of
$SU(3) \times SU(2)  \times U(1)$ where the number 
${\bf \frac{(1+
\frac{3R}{2})(2+\frac{3R}{2})}{2}}$ is
the dimension of $SU(3)$ representation while 
$\bf R+1$ is the dimension of $SU(2)$ representation.
The condition 
$\Delta=R$ saturates the bound on $\Delta$ from superconformal algebra.
The fact that the $R$-charge of a chiral operator is equal to
the dimension was observed in \cite{bhk} in the context of $R$ symmetry gauge
field. 

$1)$ One hypermultiplet:
\bea
\left(\bf \frac{(1+\frac{3R}{2})(2+\frac{3R}{2})}{2}, {\bf R+1} 
\right), \qquad
 \Delta=R.
\nonu
\eea 

According to \cite{fabbri1}, 
the information on the Laplacian eigenvalues allows us 
to get the spectrum of hypermultiplets of the theory corresponding to
the chiral operators of the SCFT.
This part of spectrum was given in \cite{fabbrietal} and the form of operators
is 
\bea
\mbox{Tr} \Phi_{\mbox{c}} \equiv \mbox{Tr} (U^3 V^2)^{R/2}
\label{chiral}
\eea
where the flavor $SU(3)$ and $SU(2)$ indices 
are totally symmetrized and the chiral superfield
$\Phi_{\mbox{c}}(x, \theta)$ satifies $D_{\al}^{+} 
\Phi_{\mbox{c}}(x, \theta) =0$. 
The hypermultiplet spectrum in the KK harmonic expansions on $M^{1,1,1}$
agrees with the chiral superfield predicted by the 
conformal gauge theory.
From this, the dimension of $U^3 V^2$ should be 2 to match the spectrum.
In fact, the conformal weight of a product of
chiral fields equals the sum of the weights of the single components.
This is due to the the relation of $\Delta=R$ satisfied by
chiral superfields and to the additivity of the $R$-charge.

$2)$ One short graviton multiplet:
$
\left(\bf \frac{(1+\frac{3R}{2})(2+\frac{3R}{2})}{2}, 
{\bf R+1} \right), \qquad
 \Delta=R+2$

The gauge theory interpretation of this multiplet is obtained by
adding a dimension 2 singlet operator with respect to 
flavor group into the above chiral superfield 
$ \Phi_{\mbox{c}}(x, \theta)$.
We consider
$
\mbox{Tr} \Phi_{\al \be} \equiv \mbox{Tr} \left( T_{\al \be}  \Phi_{\mbox{c}} 
\right),$
where $T_{\al \be}(x, \theta)$ 
is a stress energy tensor 
and $ \Phi_{\mbox{c}}(x, \theta)$ is a chiral superfield (\ref{chiral}). 
All color indices are symmetrized before taking the contraction.
This composite operator satisfies $D^{+}_{\al} 
\Phi^{\al \be}(x, \theta) =0 $.

$3)$ One short vector multiplet:
\bea
\left(\bf \frac{(1+\frac{3R}{2})(2+\frac{3R}{2})}{2}, 
{\bf R+3} \right) \;\;\; \mbox{or} \;\;\;
\left(\bf (2+\frac{3R}{2})(4+\frac{3R}{2}), 
{\bf R+1} \right), \qquad
 \Delta=R+1.
\nonu 
\eea
 
One can construct the following gauge theory object, corresponding to
the short vector multiplet
$\left(\bf \frac{(1+\frac{3R}{2})(2+\frac{3R}{2})}{2}, 
{\bf R+3} \right)$,
$
\mbox{Tr} \Phi  \equiv \mbox{Tr} \left( J_{SU(2)}  \Phi_{\mbox{c}} 
\right),$
where $J_{SU(2)}(x, \theta)$ 
is a conserved vector current transforming in the adjoint representation
of $SU(2)$ flavor group and 
$ \Phi_{\mbox{c}}(x, \theta)$ is a chiral superfield (\ref{chiral}). 
There exists
other short multiplet by $
\left(\bf (2+\frac{3R}{2})(4+\frac{3R}{2}), 
{\bf R+1} \right)
$. Similarly one can consider
$
\mbox{Tr} \Phi  \equiv \mbox{Tr} \left( J_{SU(3)}  \Phi_{\mbox{c}} 
\right),$
where $J_{SU(3)}(x, \theta)$ 
is a conserved vector current transforming in the adjoint representation
of $SU(3)$ flavor group.
In this case, we have
$D^{+ \al} D^{+}_{\al} \Phi(x, \theta) =0$. \footnote{   
Of course there 
exist two short gravitino multiplets
specified by 
\bea
\left(\bf (2+\frac{3R}{2})(4+\frac{3R}{2}), 
{\bf R+2} \right), \;\;\;
\left(\bf \frac{(1+\frac{3R}{2})(2+\frac{3R}{2})}{2}, {\bf R} 
\right)
\nonu
\eea
with conformal weight, $\Delta = R-1/2$ and $R+3/2$ respectively.} 
Therefore the short $OSp(2|4)$ multiplets 1), 2) and 3) 
saturate the unitary bound and have a 
conformal dimension related to the $R$-charge and maximal spin. 

$\bullet$ {\bf Long multiplets} \cite{fabbri1}
\footnote{
The complete KK spectrum of the round $\S^7$
compactification consists of short $OSp(8|4)$ multiplets
characterized by a quantization of masses in terms of
the $SO(8)$ $R$-symmetry representation \cite{gw}.
All the KK states are BPS states and their spectrum can be
obtained by analyzing the short unitary irreducible representation of
$OSp(8|4)$. In the dual theory, all the composite primary
operators have conformal weight equal to their naive dimensions.
No anomalous dimensions are generated. However, the KK states 
with lower supersymmetry, for example, in $AdS_4 \times M^{1,1,1}$
do not fall into short multiplets of $OSp(2|4)$ and do not
necessarily have quantized masses. Their masses depend not only on
the $R$-symmetry representation but also on the gauge group 
$SU(3) \times SU(2)$ in the
supergravity side(or flavor group in the dual field theory). 
This implies that according to AdS/CFT correspondence, anomalous
dimensions are generated. }

Although the dimensions of nonchiral operators are in general irrational,
there exist special integer values of $n_i$ such that
for $M_1=n_1, M_2=n_1+3R/2$ and $J=R/2+n_2$, 
one can see the Diophantine like 
condition(See also \cite{ahnplb}), 
\bea
n_1^2-n_1 + 3 \left( n_2^2 -n_2 \right) -6 n_1 n_2=0
\label{dequation}
\eea
make $\sqrt{36+E}$ be equal to
$4R+2(2n_1 +2n_2 +3)$. 
Furthermore in order to make 
the dimension be rational(their conformal dimensions are protected), 
$45 + E/4 -6 \sqrt{36 +E}$  in (\ref{delta}) should be square of 
something. It turns out this is the case without any further restrictions on
$n_i$'s. Therefore we  have $\De=R+ n_1 +n_2$ which is $\De_{+}$ for
$\De \geq 3/2$ and $\De_{-}$ for $\De \leq 3/2$. 
This is true if we are describing states with finite $\Delta$ and $R$.
Since we are studying the scaling limit $\Delta, R \rightarrow \infty$,
we have to modify the above analysis.
This constraint (\ref{dequation}) comes from the fact that the 
energy eigenvalue of the Laplacian on $M^{1,1,1}$ for the supergravity mode
(\ref{energy})
takes the form 
\bea
E = \frac{64}{3} n_1^2 +32 n_2^2 + \frac{64}{3} \left( \frac{3}{2} R+2 
\right)n_1 + 
32 \left(R+1 \right)n_2 + 16 R \left( R + 3 \right).    
\label{energyr}
\eea
One can show that the conformal weight of the long vector multiplet $A$
below becomes rational if the condition (\ref{dequation}) is satisfied.  

$1)$ One long vector multiplet $A$:
\bea
\left( \bf \frac{(1+n_1)(1+n_1+\frac{3R}{2})(2+ 2n_1+
\frac{3R}{2})}{2}, 
{\bf 2 n_2 + R+1} \right), 
\qquad
\Delta= -\frac{3}{2} +\frac{1}{4} \sqrt{E+36}.
\nonu 
\eea

However, as we take the limit of $R \rightarrow \infty$, 
this constraint (\ref{dequation}) is relaxed. The combination of 
$\Delta-R$ is given by   
\bea
\Delta-R = n_1 +n_2  + {\cal O}(\frac{1}{R})
\label{lowest}
\eea
where the right hand side is definitely rational and they are integers.
So the constraint  (\ref{dequation}) is not relevant in the
subsector of the Hilbert space we are interested in. 
Candidates for such states in the gauge theory side are given in terms of
semi-conserved superfields \cite{fabbrietal}. 
Although they are not chiral primaries, their
conformal dimensions are protected. The ones we are interested in take the
following form,
\bea
\mbox{Tr} \Phi_{\mbox{s.c.}} \equiv
\mbox{Tr} \left[  \left( J_{SU(3)} \right)^{n_1} \left( J_{SU(2)} 
\right)^{n_2} 
  \left(U^3 V^2 \right)^{R/2} \right]
\label{semi}
\eea   
where the scalar superfields $J_{SU(3)}(x,\theta)$ transform 
in the adjoint representation of flavor group $SU(3)$ and satisfy
$D^{\pm \al} D^{\pm}_{\al} J_{SU(3)}(x,\theta) =
0 $ with conformal dimension 1
and zero $U(1)_R$ charge. 
Similarly, the scalar superfields $J_{SU(2)}(x,\theta)$
transform in the adjoint representation of the flavor group
$SU(2)$.
Also we have $D^{+ \al} D^{+}_{\al} \Phi{
\mbox{s.c.}}(x, \theta)=0$. 
Since the singleton superfields $U_{i, bd}^{ac}$
carry  indices $a,c$ in the $\Box\!\Box$ of $SU(N)$ and 
indices $b,d$ in the ${\Box\!\Box}^\ast$ of the $SU(N)$,
the fields $V_{A,ace }^{bdf}$
carry indices $a,c,e$ in the $ {\Box\!\Box\!\Box}^\ast$ of $SU(N)$ and 
indices $b,d,f$ in the  ${\Box\!\Box\!\Box}$ of the $SU(N)$,
one can construct the following conserved flavor
currents transforming
$({\bf 8},{\bf 1})$ and $ ({\bf 1},{\bf 3})$ under
$SU(3) \times SU(2)$ respectively 
\bea
 \left( J_{SU(3)} \right)_
{i_1}^{j_1} & = & U^{j_1} \overline{U}_{i_1} -\frac{\delta_
{i_1}^{j_1}}{3} U \overline{U},\nonu \\ 
 \left( J_{SU(2)} \right)_
{i_2}^{j_2} & = & V^{j_2} \overline{V}_{i_2} -\frac{\delta_
{i_2}^{j_2}}{2} V \overline{V}, 
\nonu 
\eea
where the color indices are contracted in the right hand side.
Note that the conformal dimension of these currents is not
the one of naive sum of $U$ and $\overline{U}$ and 
$V$ and $\overline{V}$. 
As we discussed in the last section, supergravity theory in $AdS_4 \times
M^{1,1,1}$ acquires an enhanced ${\cal N}=8$ superconformal symmetry in the
Penrose limit. This implies that the spectrum of the gauge theory 
operators in this subsector should fall into
${\cal N}=8$ multiplets. We expect that both the chiral primary
fields of the form  $\mbox{Tr} (U^3 V^2)^{R/2}$ (\ref{chiral}) and the 
semi-conserved multiplets
of the form (\ref{semi}) combine into make ${\cal N}=8$ multiplets in the 
limit. Note that for finite $R$, the semi-conserved multiplets
should obey the Diophantine constraint (\ref{dequation}) in order for them to
possess rational conformal weights. 

In the remaining multiplets we consider the following particular
representations in the global symmetry group:
$
\left( \bf \frac{(1+n_1)(1+n_1+\frac{3R}{2})(2+ 2n_1+
\frac{3R}{2})}{2}, 
{\bf 2 n_2 + R+1} \right).
$

$2)$ One long graviton multiplet $h$:
$
\Delta= \frac{1}{2} +\frac{1}{4} \sqrt{E+36}
$

For finite $R$ with rational dimension, 
after inserting the $E$ into the above, 
we will arrive at the relation with same constraint (\ref{dequation})
which is greater than (\ref{lowest}) by 2:
\bea
\Delta-R = 2+ n_1 +n_2  +{\cal O} \left(\frac{1}{R} \right).
\label{delta2}
\eea
The gauge theory interpretation of this multiplet is
quite simple. If we take  a semi-conserved current 
$\Phi_{\mbox{s.c.}}(x, \theta)$ defined in (\ref{semi})
and multiply it by a stress-energy tensor superfield 
$T_{\al \be}(x, \theta)$ that is a singlet with respect to the flavor group,
namely 
$
\mbox{Tr} \left( T_{\al \be}  \Phi_{\mbox{s.c.}} 
\right)$,
we reproduce the right $OSp(2|4) \times SU(3) \times SU(2)$
representations of the long graviton multiplet.
Also one can expect that other candidate for this multiplet with different 
representation by multiplying a semi-conserved current into a quadratic
 conserved scalar superfield: 
$\mbox{Tr} \left( J_{SU(3)} J_{SU(3)}  
\Phi_{\mbox{s.c.}} \right)$, $\mbox{Tr} \left( J_{SU(3)} J_{SU(2)}  
\Phi_{\mbox{s.c.}} \right)$ or $\mbox{Tr} \left( J_{SU(2)} J_{SU(2)}  
\Phi_{\mbox{s.c.}} \right)$. In this case,
the constraint for finite $\Delta$ and $R$ is
shifted as $n_1 \rightarrow n_1+2$, $n_1 \rightarrow 
n_1 +1, n_2 \rightarrow n_2 +1$ and $n_2 \rightarrow n_2 +2$ respectively.

$3)$ One long vector multiplet $Z$:
$
\Delta= \frac{1}{2} +\frac{1}{4} \sqrt{E+32R+36}
$

Although there exists no rational dimension for this case with any choice of
$n_i$'s when $\Delta$ and $R$ are finite, 
the combination of $\Delta -R$ with Penrose limit 
$R \rightarrow \infty$ in the gauge theory side becomes
\bea
\Delta-R = 3 + n_1 +n_2  +{\cal O} \left(\frac{1}{R} \right).
\nonu
\eea
Since we do not have any singlet of conformal dimension 3  
with respect to the flavor group, one cannot increase a conformal dimension
by simply tensoring any extra superfields into a semi-conserved current
in order to match the spectrum.
So the only way to do this is to increase the number of 
conserved scalar superfield. 
In order to produce the following gauge theory operator 
$
 \mbox{Tr} \left(
T_{\al \be} J_{SU(3)} \Phi_{\mbox{s.c.}} \right)$ or
$
 \mbox{Tr} \left(
T_{\al \be} J_{SU(2)} \Phi_{\mbox{s.c.}} \right)$
corresponding to
this vector multiplet, one can think of a 
higher dimensional representation in the global symmetry $SU(3)$ or $SU(2)$. 
Then the constraint coming
from the requirement of rationality of conformal dimension is also changed
for finite $\Delta$ and $R$. One can describe also the product of
cubic $J$'s and $\Phi_{\mbox{s.c.}}$ similarly.

$4)$ One long vector multiplet  $W$:
$
\Delta= \frac{5}{2} +\frac{1}{4} \sqrt{E+36}
$

In this case, we get $\Delta -R$ by adding 2 to the one in (\ref{delta2})
\bea
\Delta-R = 4+ n_1 +n_2  +{\cal O} \left(\frac{1}{R} \right).
\nonu
\eea
One can describe corresponding gauge theory operator by taking
quadratic stress-energy tensor $ T_{\al \be} T^{\al \be}(x, \theta)$
and mutiplying it into a semi-conserved 
current $\Phi_{\mbox{s.c.}}(x, \theta)$
in order to match with 
 the conformal dimension. That is, one obtains  
$
\mbox{Tr} \left( T_{\al \be} T^{\al \be}  
\Phi_{\mbox{s.c.}} \right)$.
Similarly one can construct the following gauge theory operators
related to this vector multiplet 
$\mbox{Tr} \left(T_{\al \be} J_{SU(3)} J_{SU(3)}  
\Phi_{\mbox{s.c.}} \right)$,
$\mbox{Tr} \left(T_{\al \be} J_{SU(3)} J_{SU(2)}  
\Phi_{\mbox{s.c.}} \right)$ or 
$\mbox{Tr} \left(T_{\al \be} J_{SU(2)} J_{SU(2)}
\Phi_{\mbox{s.c.}} \right)$.
For four $J$'s with $\Phi_{\mbox{s.c.}}$,
one can analyze similarly.  

$5)$ One long vector multiplet $Z$:
$\Delta= \frac{1}{2} +\frac{1}{4} \sqrt{E+4}
$

Although there exists no rational dimension for this case with any choice of
$n_i$'s when $\Delta$ and $R$ are finite, 
the combination of $\Delta -R$ with Penrose limit 
$R \rightarrow \infty$ in the gauge theory side becomes
$
\Delta-R = 2+ n_1 +n_2  +{\cal O} \left(\frac{1}{R} \right).
$  
In addition to the above 1), 2), 3), 4) and 5) multiplets,
there are also 
six long gravitino multiplets. \footnote{ 
There exist three of them \cite{fabbri1} $\chi^{+}$
characterized by
$
\Delta  =  -\frac{1}{2} +\frac{1}{4} \sqrt{E + 16R+32},
\;\;\; \mbox{or} \;\;\; -\frac{1}{2} +\frac{1}{4} \sqrt{E + 16R}
$
and  three of them $\chi^{-}$ characterized by 
$
\Delta  =  \frac{3}{2} +\frac{1}{4} \sqrt{E + 16R+32},
\;\;\;  \mbox{or} \;\;\; \frac{3}{2} +\frac{1}{4} \sqrt{E + 16R}
$. 
Similar anaylsis can be done in this case. Although for finite 
$\Delta$ and $R$, both do not provide rational conformal dimensions,
in the Penrose limit there is no constraint on the integer values and
$R \rightarrow \infty$ limit will give us a rational conformal dimension.} 

%%%%%%%%%%%%%%%%%%%%%%%%%%%%%%%%%%%%%%%%%%%%%%%%%%%%%%%%%%%%%%%%%%%%%%%%%%%%%
\section{Conclusion}
%%%%%%%%%%%%%%%%%%%%%%%%%%%%%%%%%%%%%%%%%%%%%%%%%%%%%%%%%%%%%%%%%%%%%%%%%%%%%

We described an explicit example of an ${\cal N}=2$
superconformal field theory that has a subsector of the Hilbert space
with enhanced ${\cal N}=8$ superconformal symmetry, in the large $N$ limit
from the study of $AdS_4 \times M^{1,1,1}$.
The pp-wave geometry in the scaling limit produced to 
the maximally ${\cal N}=8$  supersymmetric pp-wave
solution of $AdS_4 \times \S^7$.
The result of this paper shares common characteristic feature 
of previous case of $AdS_4 \times Q^{1,1,1}$ \cite{ahn02}.
This subsector of gauge theory is achieved by Penrose limit
which  constrains strictly the states of the gauge theory to those
whose conformal dimension and $R$ charge diverge in the large $N$ limit
but possesses finite value $\Delta-R$.
We predicted for the spectrum of $\Delta-R$ of the ${\cal N}=2$
superconformal field theory and proposed how the exicited states in the
supergravity correspond to
gauge theory operators. In particular, both the chiral multiplets 
(\ref{chiral}) and
semi-conserved multiplets (\ref{semi}) of ${\cal N}=2$ supersymmetry should
combine into ${\cal N}=8$ chiral multiplets. 
      
\vskip1cm
$\bf Acknowledgements$

This research was supported 
by 
grant 2000-1-11200-001-3 from the Basic Research Program of the Korea
Science $\&$ Engineering Foundation.

\end{document}